\documentstyle[12pt]{article}
\input epsf.tex
\begin{document}
\begin{center}
{\Large Abelian Landau-Pomeranchuk-Migdal Effects}        

\bigskip

{\large I.M. Dremin $^1$ and C.S. Lam $^2$}       

\bigskip

$^1$Lebedev Physical Institute, Moscow 117924, Russia\\
$^2$Department of Physics, McGill University, 
Montreal, Canada H3A 2T8
\end{center}

\begin{abstract}
It is shown that the high-energy expansion of the scattering amplitude 
calculated from Feynman diagrams factorizes in such a way that it can be 
reduced to the eikonalized form up to the terms of inverse power in energy
in accordance with results obtained by solving the Klein-Gordon equation.
Therefore the two approaches when applied to the suppression of the emission 
of soft photons by fast charged particles in dense matter should give rise
to the same results. A particular limit of thin targets is briefly discussed.
\end{abstract}

\def \dyadic#1{\vbox{\ialign{##\crcr
     $\hfil
{\thinspace\scriptstyle\leftrightarrow}
\hfil$\crcr\noalign{\kern-.01pt\nointerlineskip}
     $\hfil\displaystyle{#1}\hfil$\crcr}}}
\def\C{{\cal C}}
\def\e{\epsilon}
\def\be{\begin{eqnarray}}
\def\ee{\end{eqnarray}}
\def\nn{\nonumber}
\def\({\left(}
\def\[{\left[}
\def\){\right)}
\def\]{\right]}
\def\t{\tau}
\def\p{\partial}
\def\bk#1{\langle#1\rangle}
\def\h{{1\over 2}}
\def\.{\cdot}
\def\labels#1{\label{#1}}
\def\x{\vec x}
\def\k{\vec k}
\def\b{\vec b}
\def\U{{\cal U}}
\def\R{{\cal R}}
\def\i{\infty}
\def\v#1{\vec #1}
\def\b#1{{\bf #1}}
\def\'{\,'}
\section{Introduction}
In 1953, Ter-Mikaelian \cite{ter} noticed that at high energies the 
longitudinal momentum transferred at each scattering to an electron 
traversing a medium becomes very small. It enlarges the effective formation 
length for emitted photons and leads to specific effects in crystals.
Soon, Landau and Pomeranchuk \cite{lapo} remarked that it would result
in the suppression of radiation in amorphous media if the formation length
becomes large relative to the scattering mean free path of the electron. 
Nowadays, these effects are confirmed in experiment both for crystals and for 
amorphous media and are widely discussed. 

Several theoretical approaches extending the classical
treatments of Refs \cite{ter, lapo} have been proposed. Migdal \cite{mig}
has used the Focker-Planck equation and quantified the results for amorhous 
media so that the corresponding effect is now known as Landau-Pomeranchuk-
Migdal (LPM) effect. The polarization of the medium suppresses the soft photon
emission as well (Ter-Mikaelian effect, see \cite{MTM}).
The Kharkov group \cite{khar} has applied the path-integral
technique to treat both effects simultaneously. In connection with recent 
SLAC experiments \cite{ANT} on LPM effect, Blankenbecler and Drell \cite{BD}
have proposed to use at high energies the solution of the Klein-Gordon 
equation with account of higher order corrections in expansion of the phase of 
the wave function in inverse powers of the initial momentum. The purely
diagrammatic approach has been advocated in the series of papers done in Orsay
\cite{orsa} which extended the treatment of LPM effect to the non-Abelian
case as well, in search for analogous effects in QCD started by Gyulassy and
Wang \cite{gywa}. Somewhat special treatment using the Schr\"{o}dinger equation 
and considering compact quark-antiquark-gluon systems
has been proposed by B. Zakharov \cite{zakh} who criticized some approximations
used in papers \cite{orsa}. Soon it was shown \cite{BDMS} that with account of
some additional terms omitted in \cite{orsa} the two approaches are equivalent.
Recently, Novosibirsk group \cite{baka} carefully considered different limiting
cases of LPM effect in QED confronting them to experiment \cite{ANT}.
Even though being common in spirit and close in final results, these approaches 
use different technique and different models of a medium so that sometimes it is 
hard to judge the correspondence between them. At the same time, some limiting
cases are preferable to treat by either one or another method.

Here, we would like to fill in one link in this chain of proposals and to show
that for the Abelian case the solutions of the Klein-Gordon equation obtained
in Ref. \cite{BD} directly correspond to results of summing up the series
of Feynman graphs considered in Ref. \cite{orsa} in the high energy ($p 
\rightarrow \infty $) limit i.e. up to the terms of the order of $O(p^{-1})$.
In doing this, we show that the problem can be stated in terms of the 
post-eikonal approximation because of factorization of sums of Feynman diagrams
in this limit. To simplify the formulas, we consider charged scalar particles
since spin effects can be incorporated in a straightforward manner (see 
\cite{BD}).

\section{High-Energy Wave Function in the Post-Eikonal Approximation}

Let $p^\mu=(p^0,\v p)=(\sqrt{p^2+m^2},\b 0,p)$ 
be the incoming 
four-momentum of a charged scalar particle, and 
\be
\phi(x)=e^{ip\. x}\int d^3qe^{i\v q\.\v x}
\tilde\phi(\v p;\v q)\labels{wf}\ee
its energy eigenfunction with energy $p^0$
in the presence of a static source
\be
A^0(\x)=V(\x)=\int d^3ke^{i\v k\.\v x}v(\v k),\labels{v}\ee
which transfers a total amount of momentum $\v q$
to the particle. We assume the momentum $\v k=(\b k,k_z)$
 provided by the source 
at each interaction to be
much less than the incoming momentum $p$ (more exactly,
$\v k^2\ll 2\v p\.\v k$), so that the post-eikonal
approximation
\be
{1\over (p+k)^2-m^2+i\e}=-{1\over 2\v p\.\v k+\v k^2-i\e}
\simeq -{1\over 2p}\({1\over k_z-i\e}-{\v k^2\over
 2p}{1\over (k_z-i\e)^2}\)
\labels{prop}\ee
can be applied to the particle propagators of the Feynman diagrams.
The first term in the expansion constitutes the familiar eikonal
approximation \cite{EIK}, but the second term is also needed
to describe the Landau-Pomeranchuk-Migdal (LPM) 
effect. Our aim is to calculate
$\phi(x)$ in perturbation theory to an accuracy of order of $1/p$ in all orders.


\begin{figure}
\vskip -26 cm
\centerline{\epsfxsize 7 truein \epsfbox {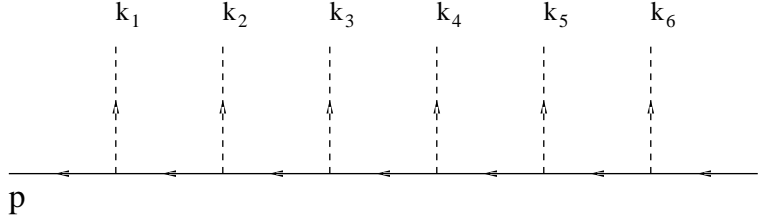}}
\nobreak
\vskip -9 cm\nobreak
\caption{A Feynman
tree diagram for the emission of $n$ photons from an energetic particle
of momentum $p$}
\end{figure}

The triple interaction vertex 
is simply $2pv(\v k)$ in
momentum space, and the seagull
vertex $\h v(\v k_1)v(\v k_2)$ is down by a factor
of $p$ compared to the triple interaction. The amplitude
in the tree approximation (see Fig.~1) is given by
\be
\tilde \phi(\v p;\v q)&=&
\sum_{n=0}^\infty\tilde\phi_n(\v p;\v q),\nn\\
\tilde\phi_n(\v p;\v q)&=&\int\(\prod_{i=1}^n d^3k_i\ v(\v k_i)\)
P_n(p;k_1,\cdots,k_n)\delta\(\sum_{i=1}^n\v k_i-\v q\),\nn\\
P_n(p;k_1,\cdots,k_n)&=&\prod_{i=1}^n{2p\over (p+K_i)^2-m^2+i\e}+\cdots,\nn\\
K_i&\equiv&\sum_{j=1}^ik_j,\nn\\
\labels{tree}\ee
where the ellipses represent seagull contributions, and $\tilde
\phi_0(\v p;\v q)\equiv 1$.
Note that the repeated
 appearance of $v(\v k)$'s in (\ref{tree})
is a result of multiple interactions with the same source $v$;
it does not necessarily imply the presence of $n$ distinct scatterers, though that can be accommodated.
Note also that $P(p;\k_1,\cdots,\k_n)$ is not symmetric in its variables $k_i$.

The perturbative expression for the wave function (\ref{wf}) is 
\be
\phi(x)&=&\sum_{n=0}^\infty e^{ip\. x}
\phi_n(\v x),\nn\\
\phi_n(\v x)&=&\int\prod_{i=1}^n\(d^3k_i\ e^{i\v k_i\.\v x}v(\v k_i)\)
P_n(p;k_1,\cdots,k_n).\labels{phin}\ee

In the eikonal approximation where the $O(\v k^2/2p)$ terms in 
(\ref{prop}) are neglected, $P_n$ becomes
\be
P_n^{(0)}(p;k_1,\cdots,k_n)=(-)^n\prod_{i=1}^n
{1\over K_{iz}-i\e}.\labels{pn0}\ee
If the correction terms in (\ref{prop}) are included to compute
$P_n=P_n^{(0)}+P_n^{(1)}/p+O(1/p^2)$, then the first post-eikonal 
contribution is
\be
P_n^{(1)}(p;k_1,\cdots,k_n)&=&-P_n^{(0)}(p;k_1,\cdots,k_n)\h
\sum_{i=1}^n{\v K_i^2\over K_{iz}-i\e}
+\cdots.
\labels{pn1}\ee
The corresponding contributions to $\phi_n$ will be denoted by
$\phi_n^{(0)}$ and $\phi_n^{(1)}$ respectively.

We shall use $\bk{f(k_1,k_2,\cdots,k_n)}$ to represent the permutation
average of any function $f$. Thus, for example, $\bk{f(k_1,k_2,k_3)}
=(f(k_1,k_2,k_3)+f(k_1,k_3,k_2)+f(k_2,k_1,k_3)+f(k_2,k_3,k_1)+
f(k_3,k_1,k_2)+f(k_3,k_2,k_1))/3!$. We may and shall
replace $P_n(k_1,\cdots,k_n)$ in (\ref{phin}) by its symmetric form
$\bk{P_n(k_1,\cdots,k_n)}$, because this allows us to
use the eikonal factorization formula \cite{EIK} (see also Appendix A)
\be
\bk{\prod_{i=1}^n{1\over K_{iz}-i\e}}={1\over n!}
\prod_{i=1}^n{1\over k_{iz}-i\e}\labels{eik}\ee
to compute the eikonal wave function:
\be
\phi^{(0)}_n(\v x)&=&{1\over n!}\(-i\chi_0(\v x)\)^n,\nn\\
\chi_0(\v x)&=&\chi_0(z,\b b)=-i\int d^3ke^{i\v k\.\v x}
v(\v k){1\over k_z-i\e}=\int_{-\i}^zV(z',\b b)dz',
\nn\\
\phi^{(0)}(x)&=&
\exp\[-ip^0x^0+ipz-i\chi_0(\v x)\].\labels{eikphi}\ee
When the post-eikonal terms in (3) are included,
it is no longer clear how factorization and
summation can be carried out. 
In order to get an idea how the post-eikonal contributions 
can be organized to yield factorization,
let us look at the second-order contribution to $\phi^{(1)}(\v x)$:
\be
&&\hskip1cm\phi^{(1)}_2(\v x)=
\int d^3k_1d^3k_2e^{i(\v k_1+\v k_2)\.\v x}
v(\v k_1)v(\v k_2)\bk{P^{(1)}_2(p;k_1,k_2)},\nn\\
&&\hskip-.5cm\bk{P^{(1)}_2(p;k_1,k_2)}=-{1\over 4}
\biggl(
{\v k_1^2\over (k_{1z}-i\e)^2}
{1\over k_{1z}+k_{2z}-i\e}
+{1\over k_{1z}-i\e}{(\v k_1+\v k_2)^2\over(k_{1z}+k_{2z}-i\e)^2}\nn\\
&&\hskip1.2cm+{\v k_2^2\over (k_{2z}-i\e)^2}
{1\over k_{1z}+k_{2z}-i\e}
+{1\over k_{2z}-i\e}{(\v k_1+\v k_2)^2\over(k_{1z}+k_{2z}-i\e)^2}
\biggr)\nn\\
&&\hskip1.2cm+\h{1\over k_{1z}+k_{2z}-i\e}.\labels{phi12}\ee
The last term comes from the seagull diagram and the first
four terms come from the post-eikonal contribution of the 
$n!=2!$ diagrams of the type shown in Fig.~1.
The coefficient for $\v k_1^2$ is 
\be
&&-{1\over 4}{1\over k_{1z}+k_{2z}-i\e}\biggl(
{1\over (k_{1z}-i\e)^2}+
\[{1\over k_{1z}-i\e}+{1\over k_{2z}-i\e}\]{1\over k_{1z}+k_{2z}-i\e}\biggr)\nn\\
&&=-{1\over 4}{1\over (k_{1z}-i\e)^2}{1\over k_{2z}-i\e}.\labels{k12}\ee
Similarly, the $\v k_2^2$ coefficient is 
$-(k_{2z}-i\e)^{-2}(k_{1z}-i\e)^{-1}/4$, and the $\v k_1\.\v k_2$
coefficient is $-[(k_{1z}-i\e)(k_{2z}-i\e)(k_{1z}+k_{2z}-i\e)]^{-1}/2$.
Combining this last coefficient with the seagull term in (\ref{phi12})
reduces $\v k_1\.\v k_2$ to the transverse dot product 
$\b k_1\.\b k_2$. In short, the rather complicated expression in
(\ref{phi12}) can be reduced to the vastly simpler form
\be
\bk{P_2^{(1)}(p;k_1,k_2)}&=&-{1\over 2!}\h
{1\over k_{1z}-i\e}{1\over k_{2z}-i\e}
\biggl({\v k_1^2\over k_{1z}-i\e}+{\v k_2^2\over k_{2z}-i\e}
\nn\\
&+&2{\b k_1\.\b k_2\over k_{1z}+k_{2z}-i\e}\biggr).\labels{p12}\ee
Moreover, it is shown in Appendix A that 
this simplification occurs at all $n$, so
that
\be
\bk{P_n^{(1)}(p;k_1,\cdots,k_n)}&=&(-)^{n-1}{1\over n!}\h
\prod_{i=1}^n{1\over k_{iz}-i\e}
\biggl(\sum_{i=1}^n{\v k_i^2\over k_{iz}-i\e}\nn\\
&+&2\sum_{i>j}{\b k_i\.\b k_j\over k_{iz}+k_{jz}-i\e}\biggr).\labels{p1n}\ee
With this formula, the post-eikonal component of the wave
function can be factorized into
\be
\phi^{(1)}_n(\v x)&=&\phi_n^{(1)}(z,\b b)=
{1\over (n-1)!}(-i\chi_0)^{n-1}\chi_2
+{1\over (n-2)!}(-i\chi_0)^{n-2}(-i\chi_1),\nn\\
\chi_2(z,\b b)&=&
\h\int d^3ke^{i\v k\.\v x}v(\v k){\v k^{2}\over (k_z-i\e)^2}=\h\int_{-\i}^zdz'\nabla^2\chi_0(z',\b b),\nn\\
\chi_1(z,\b b)&=&-{i\over 2}
\int d^3kd^3k'e^{i(\k+\k')\.\x}v(\k)v(\k'){\b k\.\b k'
\over (k_z-i\e)(k'_z-i\e)(k_z+k'_z-i\e)}\nn\\
&=&\h\int_{-\i}^zdz'\(\b\nabla_\perp\chi_0(z',\b b)\)^2.\labels{phi1n}\ee
This factorization allows the sum over $n$
to be carried out to yield, to accuracy $O(p^{-1})$,
\be
\phi(x)&=&\exp\[-ip^0x^0+ipz-i\chi_0(\v x)-i{1\over p}\(\chi_1(\v x)+i\chi_2
(\v x)\)\].\labels{expphi}\ee
This expression agrees with the wave function
 obtained by Blankenbecler and Drell
\cite{BD} by solving the Klein-Gordon equation
\be
\[(E-V)^2+\nabla^2-m^2\]\phi(\v x)=0.\labels{kg}\ee
They looked for a solution of the form $\phi(\v x)=\exp[i\Phi(\v x)]$
accurate to order $1/p$ in $\Phi(\v x)$, and found (\ref{expphi}) to be the 
solution. As shown in Ref. \cite{BD} the extension to spinor particles
is straightforward.

In summary, we have demonstrated that factorization of sums of 
Feynman diagrams does occur even in the post-eikonal approximation,
in such a way to enable
 the perturbation series for the wave function to sum up to an exponential
form, a form that agrees with the one obtained directly by solving the
Klein-Gordon equation to accuracy $O(p^{-1})$ in the phase of the wave 
function.

\section{Outgoing Wave Function}
In the previous section we have computed the energy eigenfunction
$\phi(x)$ with incoming momentum $p^\mu$. In a similar way we can
compute the energy eigenfunction $\phi'(x)$ with outgoing momentum
${p\'}^\mu$:
\be
{\phi'}^*(x)&=&e^{-ip'\. x}\int d^3q'e^{i\v q\'\.\x}
\tilde{\phi'}^*(\v p\';\v q\'),\nn\\
\tilde{\phi'}^*(\v p\';\v q\')&=&
\sum_{n=0}^\infty\tilde{\phi'}^*_n(\v p\';\v q\'),\nn\\
\tilde{\phi'}^*_n(\v p\';\v q\')&=&
\int\(\prod_{i=1}^n d^3k_i'\ v(\v k\'_i)\)
P'_n(p';k_1',\cdots,k_n')\delta\(\sum_{i=1}^n\v k\'_i-\v q\'\),\nn\\
P'_n(p';k_1',\cdots,k_n')&=&\prod_{i=1}^n{2p'\over
(p'-K'_i)^2-m^2+i\e}+\cdots,\nn\\
K'_i&\equiv&\sum_{j-1}^ik'_j.\labels{phif1}\ee
Since the transverse component $\b p'$ of $\v p\'$ is not necessarily zero, the 
post-eikonal expansion (\ref{prop}) must be modified to read
\be
{1\over (p'-k')^2-m^2+i\e}&=&{1\over 2\v p\'\.\v k\'-{\v k\'}^2+i\e}\nn\\
&\simeq& {1\over 2p'}\({1\over k'_z+i\e}+{{\v k\'}^2-2\b p'\.\b k'\over
 2p'}{1\over (k'_z+i\e)^2}\).\labels{propf}\ee
Following very similar arguments as before, 
we finally obtain
\be
{\phi'}^*(x)&=&\exp\[-ip'\.x-i\chi'_0(\v x)-i{1\over p'}\(\chi'_1(\v x)+i\chi'_2
(\v x)\)\],\nn\\
\chi'_0(\v x)&=&i\int d^3ke^{i\v k\.\v x}
v(\v k){1\over k_z+i\e}=\int^{\i}_zV(z',\b b)dz',
\nn\\
\chi'_2(z,\b b)&=&
\h\int d^3ke^{i\v k\.\v x}v(\v k){\v k^{2}\over (k_z+i\e)^2}=\h\int^{\i}_zdz'\nabla^2\chi'_0(z',\b b),\nn\\
\chi'_1(z,\b b)&=&{i\over 2}
\int d^3kd^3k'e^{i(\k+\k')\.\x}v(\k)v(\k'){\b k\.\b k'
\over (k_z+i\e)(k'_z+i\e)(k_z+k'_z+i\e)}\nn\\
&+&
{i\over 2}\int d^3ke^{i\v k\.\v x}v(\v k){\b p'\.\b k\over (k_z+i\e)^2}
\nn\\
&=&\h\int^{\i}_zdz'\[\(\b\nabla_\perp\chi'_0(z',\b b)\)^2+2\b p'\.\nabla_\perp\chi'_0(z',
\b b)\].
\labels{treef}\ee
Let us note that the term linear in $v$ appears in $\chi'_1$ because the 
transverse momentum $\bf {p}'$ differs from zero while $\chi _1$ contains
quadratic in $v$ terms only.

\section{Multiple Scattering and Bremsstrahlung}
The on-shell scattering matrix element, with incoming momentum $p^\mu$
and outgoing momentum ${p'}^\mu$, is given by
\be
m_{fi}&=&\int d^4x{\phi'}^*(x)V(\x)e^{i p\.x}
=\int d^4xe^{-ip\'\.x}V(\x)\phi(\x)\nn\\
&=&\sum_{n=1}^\i\int\(\prod_{i=1}^nd^3k_iv(\v k_i)\)
P'_n(p';k_1,\cdots,k_{n})(2\pi)^4
\delta^4\(p'-p-\sum_{i=1}^nk_i\)\nn\\
&=&\sum_{n=1}^\i\int\(\prod_{i=1}^nd^3k_iv(\v k_i)\)
P_n(p;k_1,\cdots,k_{n})(2\pi)^4
\delta^4\(p'-p-\sum_{i=1}^nk_i\),
\nn\\ &&\labels{scatt}\ee
with $k^0_i=0$ because the source is static.

If, as a result of the scattering, a photon of four-momentum $r^\mu$ and polarization vector
$\varepsilon^*(r)$ is emitted
from the scalar particle, then the matrix element is
\be
M_{fi}(r)&=&-ie\int d^4x{\phi'}^*(x)\varepsilon^*(r)\.
\dyadic\p\phi(x)\nn\\
&=&\sum_{m,n=0}^\i \tilde{\phi'}^*_m(\v p\';\v q\')e\varepsilon^*(r)
\.\(2p+2\sum_{i=1}^nk_i-r\)
\tilde\phi_n(\v p;\v q)\nn\\
&&\hskip1in (2\pi)^4\delta^4\(p'+r-p-\sum_{i=1}^{m+n}k_i\).\labels{pho1}
\ee

The results obtained in the previous sections for the wave functions enable us
to claim that up to accuracy $O(p^{-1})$, the treatment of Abelian LPM effects
by Blankenbecler and Drell \cite{BD} who consider the Klein-Gordon equation
directly, corresponds to that of R. Baier et al \cite{orsa} who use the 
diagrammatic approach. At high energies and for finite targets the first 
approach can become preferable because of the usage of simplified expanded 
propagators and direct treatment of spatial evolution. 

It could become especially simple
in case of thin targets, where emission at single scattering with some 
corrections due to double scattering (see Appendix B) dominates, and formulas
for $P^{(1)}_1$ and $P^{(1)}_2$ as given by (\ref{p1n}) are exploited. The
slightly corrected Bethe-Heitler regime is at work. 

For finite targets, there are three lengths important in the problem. 
Those are the target thickness $l$, the mean free path $l_m$ and the 
formation length $l_f$. For soft photons, $l_{f}\approx 2\gamma ^{2}/\omega $.
Here $\gamma =p^{0}/m$, $\omega \equiv r^0$ is the photon energy. 
For screened Coulomb fields in QED, $l_{m}^{-1} = 4nr_{e}^{2}Z^{2}\ln (183/
Z^{1/3})$, where $n$ is the density of the scattering centers, $r_e =\alpha /m
\approx 2.8fm$. Let us consider the case of very thin targets and high 
energy electrons when $l \ll l_m \ll l_f$. We show that the 
decline from the Bethe-Heitler formula is determined by the target thickness
in units of the mean free path. 

Following Ref.~\cite{BD}, one can express 
the total intensity of the radiation $I(tot)$ as
the Bethe-Heitler
intensity $I(BH)\approx 4T$ (where $T=\pi l/3l_m$) suppressed by a form 
factor $F$:
\begin{equation}
I(tot)=I(BH)F. \label{itot}
\end{equation}
{}From formulas (9.16) of Ref.~\cite{BD}, it is easy to find out that for thin targets 
in the soft photon limit the form factor is given by
\begin{equation}
F\approx 1-1.5T. \label{for}
\end{equation}
The correction is small for $T$ small enough and vanishes linearly with 
target thickness.

At first sight, it seems that the 
small sizes of hadronic targets favor this limiting case for 
non-Abelian effects. However, this statement requires
 further study since
the nuclear target thickness, in units of the mean free path $T_{nucl}$ (which is
analogous to $T$), could be rather large.
 Then, inspired by eq. (\ref{for}) of QED, one would guess that the QCD 
suppression can become strong, i.e., nuclear LPM effect essential even though the target thickness in units of
the formation length is very small.

\section{Conclusions}

Thus we have shown the equivalence of two approaches to the problems of 
scattering and radiation of high-energy electrons traversing the dense medium.
One of them \cite{BD} deals with the solution of the Klein-Gordon equation,
and another one \cite{orsa} with sums of Feynman graphs. Even though the 
initial formulas have quite different forms, they lead to essentially the 
same expressions for the solutions of the above problems. This is due to the 
fact that the sum of Feynman diagrams calculated at $O(p^{-1})$ accuracy 
factorizes still in such a way that it can be represented by the corresponding
post-eikonal expansion of the phases of the wave functions.

Besides, we have argued that the nuclear LPM effect can be essential even
though the size of nuclear targets in metric units is very small.

\bigskip

\section*{Acknowledgements}
\bigskip

I.D. is grateful for support by McGill University, by the Russian 
Fund for Basic Research (grant 96-02-16347), and by INTAS.
The research of C.S.L. is supported by the Natural Sciences and Engineering
Research Council of Canada.

\appendix
\section{Factorization in the Post-Eikonal Approximation}
The aim of this appendix is to prove eq.~(\ref{p1n}). 
It would be useful first to review how (\ref{eik}) is arrived at.

The integral
\be
i^n\int_{R(12\cdots n)} d^nt\ \exp\(i\sum_{j=1}^nk_{jz}t_j\)
=\prod_{i=1}^n{1\over\sum_{j=1}^ik_{jz}-i\e}\labels{int}\ee
gives the left-hand side of (\ref{eik}) before taking the permutation
average, where the hyper-triangular 
integration region $R(12\cdots n)$ is defined to be
$\{0\ge t_1\ge t_2\ge\cdots\ge t_n>-\i\}$.
The permutation sum on the left of (\ref{eik}) can be obtained
by summing the integral over all permuted regions, whose union is
the hyper-rectangular
region $\{0\ge t_i>-\i\}$. Upon integration over the
hyper-rectangle and a division by
$n!$, one obtains the right-hand side of
(\ref{eik}) and hence the eikonal formula.

The $\v k_m^2$ coefficient of $P^{(1)}_n(k_1,\cdots,k_n)$
can be seen from (\ref{pn0}) and (\ref{pn1}) to be
\be
\h{\p\over\p k_{mz}}P_n^{(0)}(k_1,\cdots,k_n).\ee
Moreover, this relation persists upon permutation averaging.
Thus the $\v k_m^2$ coefficient of $\bk{P_n^{(1)}}$ can be obtained
by differentiating both sides of (\ref{eik}) with respect to $k_{mz}$.
The result is the one given in (\ref{p1n}).

It also follows from (\ref{pn0}) and (\ref{pn1}) that
the $2\v k_\ell\.\v k_m$ coefficient of $P_n^{(1)}(k_1,\cdots,k_n)$
can be obtained by differentiating $P_n^{(0)}$ with respect to $k_{pz}$,
where $p$ is the larger of the two numbers $\ell$ and $m$,
{\it viz.,} the one that stands to the right of the identity
permutation $(12\cdots n)$. Upon 
permutation, this relation still holds provided $p$ is taken to
be the number $\ell$ or $m$
 that stands to the right of the other.
The permutation sum will now be divided into two sums, one with
$\ell$ standing to the right of $m$, and the other
 with $m$ standing to the
right of $\ell$. In the first case, the total integration region
is a product of the triangular region $\{0\ge t_>m\ge t_\ell>-\i\}$
with the hyper-rectangular regions $\{0\ge t_i>-\i\}$
of the remaining $n-2$ variables. The result is
\be
\(\prod_{i\not=\ell.m}^n{1\over k_{iz}-i\e}\){1\over k_{mz}-i\e}{1\over k_{mz}+k_{\ell z}-i\e}.\labels{cross}\ee
Upon differentiation with respect to $k_{pz}=k_{\ell z}$ it produces
an additional factor of $-(k_{\ell z}+k_{mz}-i\e)^{-1}$.

In the second case, $m$ and $\ell$
are interchanged, but the differentiation with respect to $k_{pz}=k_{mz}$
still produces the same additional factor as before. Adding up these
two cases, we obtain (\ref{pn0}) multiplied the additional factor
$-k_{\ell z}+k_{mz}-i\e)^{-1}$, which is then equal to the
coefficient of $2\v k_\ell\.\v k_m$ in (\ref{p1n}). As before,
when combined with the seagull contributions, $2\v k_\ell\.\v k_{m}$
is reduced to $2\b k_{\ell}\.\b k_{m}$ as shown in (\ref{p1n}).

\section{Thin Targets}
For thin targets, the photon emission at single scattering (Bethe-Heitler 
regime) becomes prevailing, though there are
 corrections due to double-scattering
processes. We write down some formulas for this case. 

According to eqs (\ref{phi1n}), (\ref{expphi}), (\ref{treef}), 
(\ref{scatt}), and (\ref{pho1}),
the matrix element of soft photon
emission by a scalar electron, scattered 
once with longitudinal momentum transfer much less than $p$,
is given by
\begin{equation}
M_{1}\propto -i\int d^{3}x {\varepsilon } ^{*}(r) \cdot p
[\chi ^{t}_{0}+
\frac {1}{p}(\chi ^{t}_{1}+i\chi ^{t}_{2})], \label{M1}
\end{equation}
where $\chi ^{t}_{j}=\chi _{j}+\chi'_{j}$, and only 
terms linear in $v$ are kept in
$\chi _1$ and $\chi'_1$. It is easy to get
\begin{equation}
\chi _{0}^{t}=2\pi\int d^{2}{\b k}v({\b k},0)e^{i{\b k}\cdot{\b b}} \label{ch0}
\end{equation}
\begin{equation}
\chi ^{t}_{1}+i\chi ^{t}_{2}=\frac {i}{2}\int d^{3}kv(\v k)e^{i{\v k}\cdot {\v x}}
\left[ \frac {{\v k}^2}{(k_z -i\epsilon )^2}+\frac {{\v k}^2}{(k_z+i\epsilon)^2}
+\frac { {\b p}'\cdot{\b k}}{(k_z+i\epsilon )^2}\right].  \label{sum}
\end{equation}
These expressions follow directly both from Feynman diagrams and from the 
solution of the Klein-Gordon equation obtained in Ref. \cite{BD}.

The matrix element of photon emission at double scattering is given by 
eq. (\ref{pho1}) for $m+n=2$. The two terms with $m=0$ and $n=0$ describe
emission before or after the scattering, while the term with $m=n=1$ 
corresponds to photons emitted between two scatterings. It is easy to check 
that $1/p$-contribution contains the following factor in the integrand:
\begin{eqnarray}
I^{(2)}&\propto& {\b k}\cdot{\b k}'\biggl[-\frac {1}{(k_z-i\epsilon )
(k'_z-i\epsilon )
(k_z+k'_z-i\epsilon )}\nn\\
&&\hskip1cm+\frac {1}{(k_z+i\epsilon )(k'_z+i\epsilon )
(k_z+k'_z+i\epsilon )}\biggr ] \nonumber\\
&-&\pi i\delta(k_z){\v k}'^{2}\left [\frac {1}
{(k'_z-i\epsilon )^2}+\frac {1}{(k'_z+i\epsilon )^2}\right ]\nn\\
&-&\pi i\delta(k'_z){\v k}^{2}\left [\frac {1}
{(k_z-i\epsilon )^2}-\frac {1}{(k_z+i\epsilon )^2}\right ]
\nn\\
&-&\pi i\delta(k_z)\frac{\b p'\cdot\b k'}{(k'_z+i\e)^2}
-\pi i\delta(k'_z)\frac{\b p'\cdot\b k}{(k_z+i\e)^2}.
\label{I2}
\end{eqnarray}
This expression can be also obtained both directly from propagators in
Feynman graphs and from phases of the wave functions. Its structure is clear 
from above formulas.

\bigskip


\end{document}